\documentclass[12pt]{iopart}
\usepackage{latexsym}
\usepackage{epsfig}

\newcommand{\beq}{\begin{eqnarray}}
\newcommand{\eeq}{\end{eqnarray}}
\newcommand{\be}{\begin{eqnarray*}}
\newcommand{\ee}{\end{eqnarray*}}

\newcommand{\cc}{{c\bar{c}}}

\begin{document}

\title{Charmonium dynamics in $dA$ and $AA$ at RHIC and LHC}

\author{K.~Tywoniuk$^1$, L.~Bravina$^1$, A.~Capella$^2$, E.G.~Ferreiro$^3$, A.B.~Kaidalov$^4$ and E.~Zabrodin$^1$}

\address{$^1$ Department of Physics, University of Oslo, 0316 Oslo, Norway}
\address{$^2$ Laboratoire de Physique Th\'eorique\footnote{Unit\'e Mixte de Recherche UMR n$^{\circ}$ 8627 - CNRS}, Universit\'e de Paris XI, B\^atiment 210, 91405 Orsay Cedex, France}
\address{$^3$ Departamento de F{\'\i}sica de Part{\'\i}culas, Universidad de Santiago de Compostela, 15782 Santiago de Compostela, Spain}
\address{$^4$ Institute of Theoretical and Experimental Physics, RU-117259 Moscow, Russia}
\ead{konrad.tywoniuk@fys.uio.no}

\begin{abstract}
We discuss features of charmonium suppression at $\sqrt{s} = 200$ GeV within the framework of the Glauber-Gribov theory and the comovers interaction model. The latter approach has been extended by allowing for secondary charmonium production due to recombination of $\cc$ pairs in the medium, estimated from $pp$ data at the same bombarding energy. Centrality and rapidity dependence of the nuclear modification factor for $J/\psi$ in d+Au, Cu+Cu and Au+Au collisions at RHIC are reproduced without fitting a single model parameter. A strong suppression of $J/\psi$ is predicted for LHC energies.
\end{abstract}


\section{Introduction}
Charmonium production off nuclei is one of the most promising probes for studying the properties of matter created in ultrarelativistic heavy-ion collisions.
It was realized long time ago, that only collisions where a large density of particles is produced in the central region give rise to an anomalous $J/\psi$ suppression, i.e. above the one observed in $pA$ collisions. The anomalous suppression was advocated as a signal of charmonium melting in a thermalized quark-gluon plasma, 
but can also be described as a final state interaction of the $J/\psi$ with co-moving matter, of both partonic and hadronic nature, see e.g. \cite{Brodsky:1988,Koch:1990}. At RHIC, charmonium has been measured for several collision systems and both at forward and mid-rapidities, and hence provide unique insights into the production and interaction of charmonium in nuclear environments at very high energies.

We present an approach based on Glauber-Gribov theory, which encompasses several nuclear effects for $J/\psi$ suppression in $dA$ collisions, supplemented with additional interaction with comovers in the final state in $AA$, which also allows for secondary $J/\psi$ production from recombination.

\section{Baseline: charmonium in $dA$ collisions}
\begin{figure}[b]
  \begin{center}
    \vspace{-0.6cm}
    \includegraphics[scale=0.35]{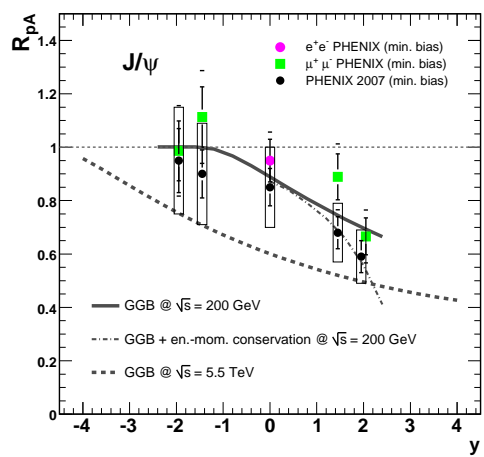}%
    \vspace{-0.4cm}
    \caption{\label{fig:dAuJpsi} Rapidity dependence of $J/\psi$ suppression for minimum bias d+Au collisions at RHIC and predictions for p+Pb collisions at LHC. Data are taken from \cite{Adler:2005ph,Adare:2007gn}.}
    \vspace{-0.6cm}
  \end{center}
\end{figure}
At lower energies, nuclear suppression of charmonium in $pA$ collisions was attributed to the successive interaction of the produced $J/\psi$ (or, rather, the pre-resonant $c\bar{c}$ pair) with the surrounding nuclear matter. In the Glauber model, this mechanism could be quantified by an absorptive cross section, $\sigma_{abs}^\psi$, which was found to be about 5 mb at $\sqrt{s} = 19$ GeV \cite{Abreu:2002rn}. Within this semiclassical picture, most models predicted a growth of absorption with energy, see e.g. \cite{Bedjidian:2003}. On the contrary, measurements of $J/\psi$ production in d+Au collisions at RHIC \cite{Adler:2005ph,Adare:2007gn} revealed a significant reduction of nuclear absorption compared to measurements at lower energies.

The RHIC results signal a breakdown of the semiclassical probabilistic picture of ordered multiple scattering, and is in line with predictions from the relativistic Glauber-Gribov theory of nuclear interactions \cite{Boreskov:1993,Braun:1998}. Above a certain critical energy $E_M$ the incoming hadron can fluctuate into a state containing the both heavy $c\bar{c}$ state and light quarks and gluons long before the collision with the nucleus takes place. In the coherent limit, both the soft partons of the fluctuation and the heavy $c\bar{c}$ system itself can interact almost simultaneously with several nucleons in the target. Schematically, the former process leads to shadowing of nuclear parton densities and dominates at mid-rapidity while the latter imposes limits from energy-momentum conservation in the forward region. In other words, the diagrams involving multiple scattering ordered in the longitudinal direction, the so-called Glauber-type diagrams, are suppressed at $E > E_M$ and hence nuclear absorption drops out.
Note, that we do not assume any diminution of the absorptive cross section.

We refer to \cite{Capella:2007,Arsene:2008} for further details on the Glauber-Gribov theory and analysis of RHIC data. In figure~\ref{fig:dAuJpsi} we compare calculations of gluon shadowing \cite{Tywoniuk:2007} alone (solid curve) and additionally with energy-momentum conservation (dash-dotted curve) to data on $J/\psi$ suppression in d+Au collisions at $\sqrt{s} = 200$ GeV \cite{Adler:2005ph,Adare:2007gn}. This constitutes the baseline in the search for the origin of anomalous suppression in $AA$ collisions.

\section{Charmonium recombination and dissociation in $AA$}
\begin{figure}[t]
  \begin{center}
    \includegraphics[width=0.45\textwidth,height=6cm]{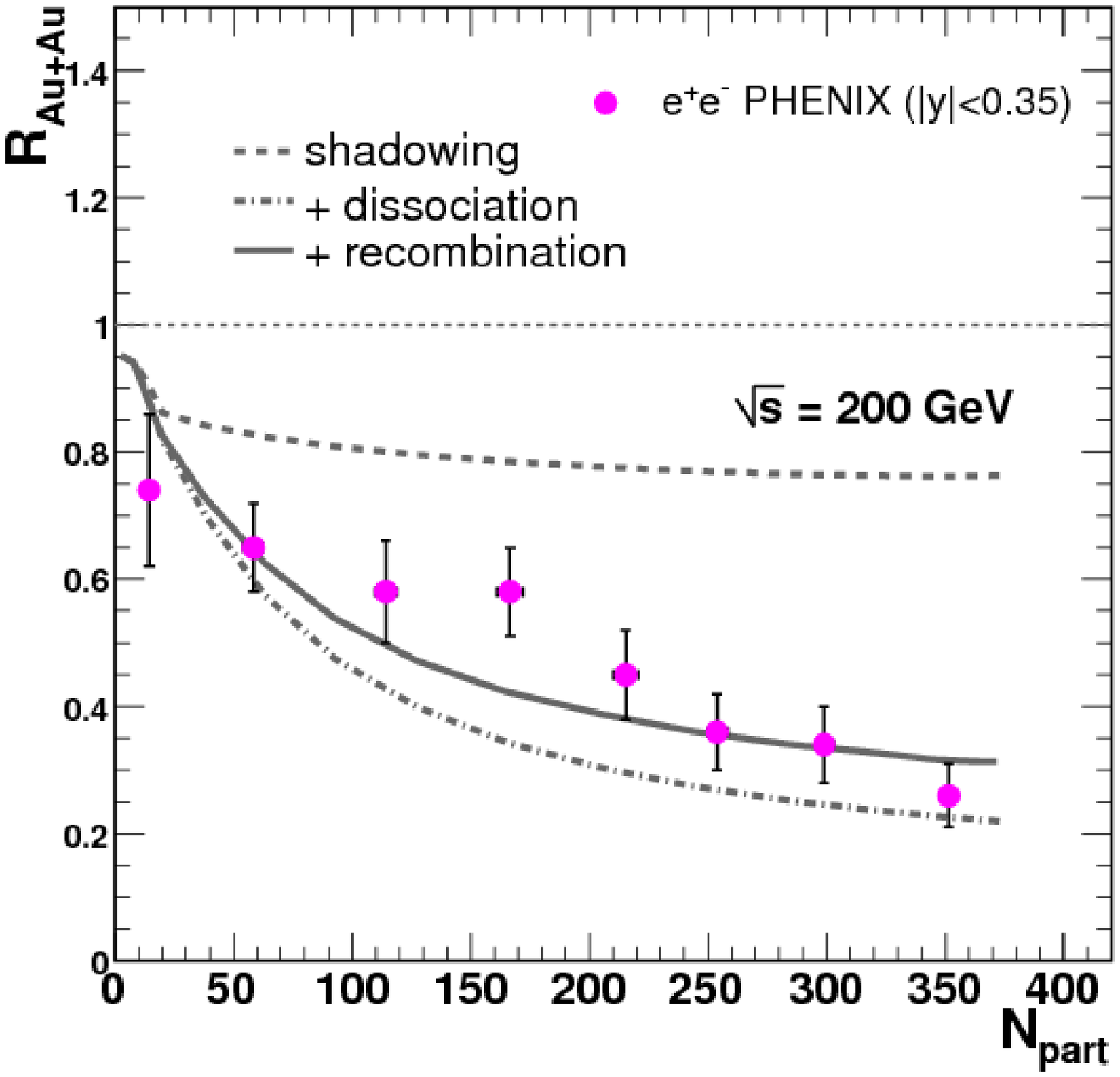}%
    \includegraphics[width=0.45\textwidth,height=6cm]{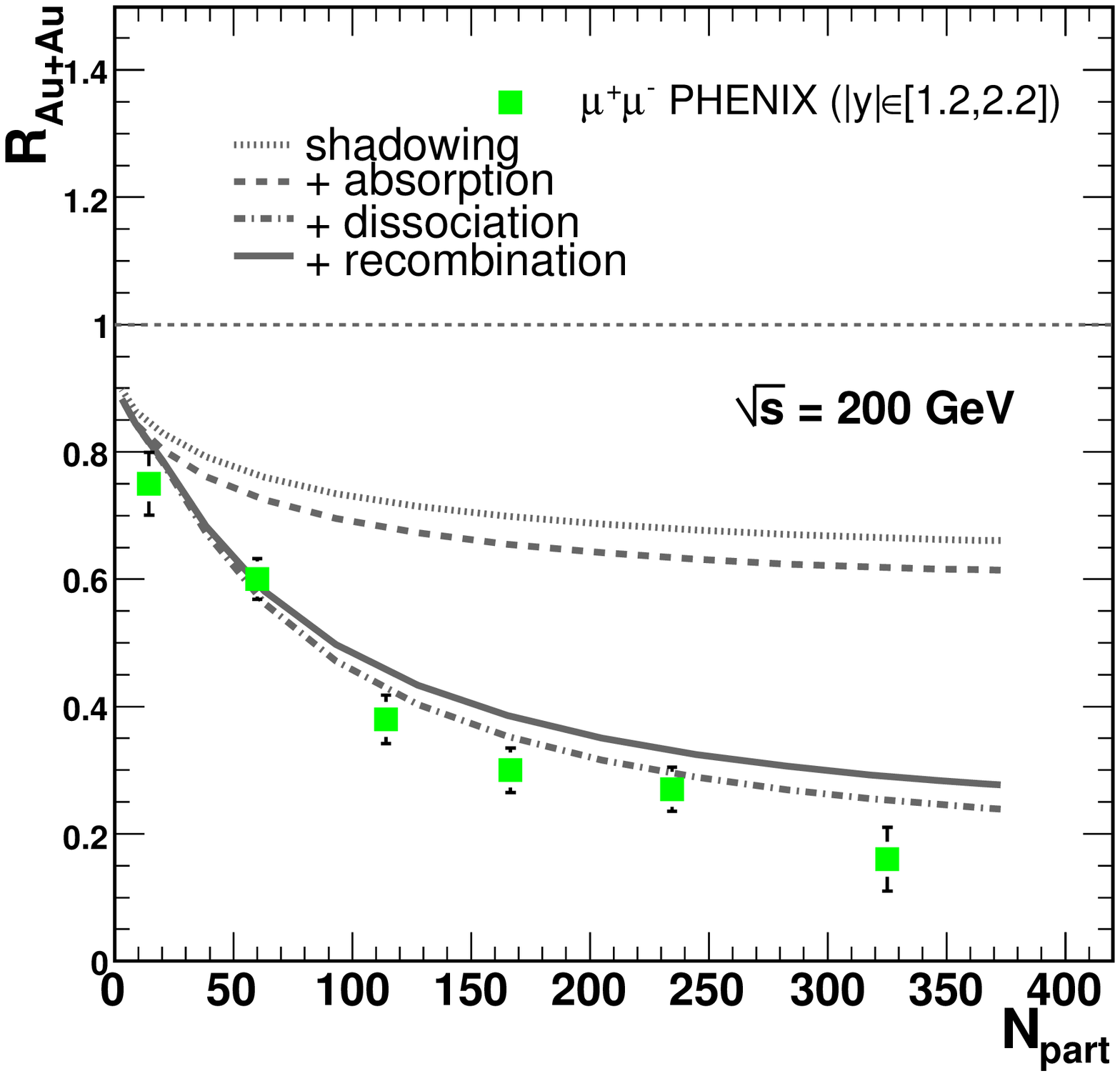}
    \vspace{-0.4cm}
    \caption{\label{fig:AuAuJpsi} Centrality dependence of $J/\psi$ suppression in Au+Au collisions at RHIC at mid- (left figure) and forward (right figure) rapidities. Data are taken from \cite{Adare:2007pr}.}
    \vspace{-0.6cm}
  \end{center}
\end{figure}
One finds a large density of particles in central $AA$ collisions at RHIC. In addition to the nuclear effects present in the smaller $pA$ collisions described above, one should also take into account the possible interaction of $J/\psi$'s with co-moving matter produced in the collision. This interaction can be quantified by a simple rate equation which, assuming boost-invariant longitudinal expansion, takes the form
\beq
\label{eq:comov}
\tau \frac{d N_{J/\psi}}{d \tau} = - \sigma_{co} \left[N^{co} N_{J/\psi} \,-\, N_c N_{\bar{c}} \right] \;,
\eeq
where $N^{co}$, $N_{J/\psi}$ and $N_{c (\bar{c})}$ denote the density of comovers, hidden and open charm at the impact parameter of the initially produced $J/\psi$, respectively. The first term in (\ref{eq:comov}) is responsible for charmonium dissociation while the second gives rise to secondary $J/\psi$ production due to recombination. The interaction cross section $\sigma_{co}$ was found at SPS to be 0.65 mb in a model where charmonium recombination was neglected due to the small density of open charm \cite{Armesto:1999}.

Calculations including both terms in (\ref{eq:comov}) and the same $\sigma_{co}$ were presented in \cite{Capella:2007aa}. The density of charmonium and open charm at RHIC was inferred from measurements in $pp$ collisions at the same energy \cite{Adare:2007,Adare:2006} (the density of open charm at forward rapidity was taken from PYTHIA). Results for $J/\psi$ suppression in Au+Au collisions at both mid- and forward rapidities ($\eta = 2.2$) at RHIC are compared to experimental data \cite{Adare:2007pr} in figure~\ref{fig:AuAuJpsi}. Note, that no parameters were fitted to obtain such good agreement with the experimental data. In particular, the stronger suppression at forward rapidities is a result of strong initial state suppression and smaller recombination. Calculations for Cu+Cu collisions at the same energy are also matching the experimental data. We refer to \cite{Capella:2007aa} for further details of the calculation.

\begin{figure}[t]
  \begin{center}
    \includegraphics[width=0.5\textwidth,height=6.5cm]{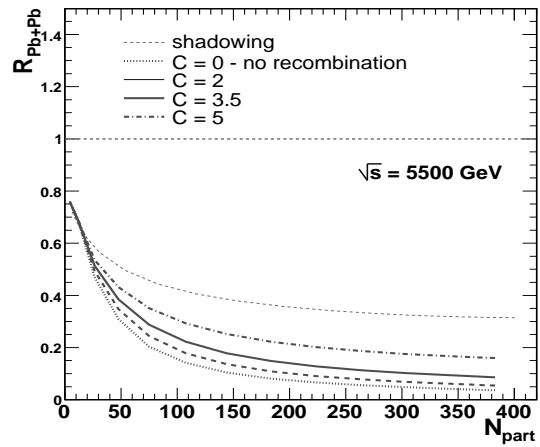}
    \vspace{-0.4cm}
    \caption{\label{fig:PbPbJpsi} Predictions for the centrality dependence of $J/\psi$ suppression at mid-rapidity in Pb+Pb collisions at LHC.}
    \vspace{-0.6cm}
  \end{center}
\end{figure}
Recombination effects will be of crucial importance in Pb+Pb collisions at $\sqrt{s} = 5.5$ TeV. Predictions for $J/\psi$ suppression at LHC have been made assuming $d \sigma_{c\bar{c}} \big/ d y \approx 1$~mb at mid-rapidity and the non-diffractive $pp$ cross section $\sigma_{pp} = 59$ mb \cite{Capella:2007aa}, with $\sigma_{co}$ kept fixed. This corresponds to $C = 2.5$ in figure~\ref{fig:PbPbJpsi}. In our model, this stands for four times stronger recombination effects at LHC than at RHIC. The strong and almost impact parameter independent $J/\psi$ suppression is mainly due to strong initial state gluon shadowing, depicted with a dotted curve in figure~\ref{fig:PbPbJpsi}, and because of a large density of comovers leading to strong dissociation.
This is in stark contrast to predictions of enhanced production of $J/\psi$ obtained in a model assuming thermalization of the heavy quarks with the entire partonic system formed at a given impact parameter \cite{Andronic:2006}.

\end{document}